\documentclass[12pt,preprint]{aastex}
\def\degpoint{\ifmmode ^{\rm{o}}\!. \else $^{\rm{o}}\!.$\fi}
\newcommand{\degrees}{$^{\rm{o}}$}
\newcommand{\ms}{\mbox{m\,s$^{-1}$}}

\newcommand{\Msun}{\mbox{M$_{\odot}$}}

\newcommand{\Mjup}{\mbox{M$_{\rm Jup}$}}

\newcommand{\Lsun}{\mbox{L$_{\odot}$}}
\newcommand{\Mearth}{\mbox{M$_{\oplus}$}}

\newcommand{\ltsimeq}{\raisebox{-0.6ex}{$\,\stackrel
         {\raisebox{-.2ex}{$\textstyle <$}}{\sim}\,$}}

\begin{document}

\title{GJ 832c: A super-earth in the habitable zone\footnote{This paper 
includes data gathered with the 6.5 meter Magellan Telescopes located at 
Las Campanas Observatory, Chile.} }

\author{Robert A.~Wittenmyer\altaffilmark{1,2,7}, Mikko 
Tuomi\altaffilmark{3,4}, R.P.~Butler\altaffilmark{5}, 
H.R.A.~Jones\altaffilmark{3}, Guillem 
Anglada-Escud{\'e}\altaffilmark{6}, Jonathan Horner\altaffilmark{7,1,2}, 
C.G.~Tinney\altaffilmark{1,2}, J.P.~Marshall\altaffilmark{1,2}, 
B.D.~Carter\altaffilmark{7}, J.~Bailey\altaffilmark{1,2}, 
G.S.~Salter\altaffilmark{1,2}, S.J.~O'Toole\altaffilmark{8}, 
D.~Wright\altaffilmark{1,2}, J.D. Crane\altaffilmark{9}, S.A. 
Schectman\altaffilmark{9}, P.~Arriagada\altaffilmark{5}, 
I.~Thompson\altaffilmark{9}, D.~Minniti\altaffilmark{10,11}, 
J.S.~Jenkins\altaffilmark{12} \& M.~Diaz\altaffilmark{12} }

\altaffiltext{1}{School of Physics, UNSW Australia, Sydney 2052, 
Australia}
\altaffiltext{2}{Australian Centre for Astrobiology, UNSW Australia, 
Sydney 2052, Australia}
\altaffiltext{3}{University of Hertfordshire, Centre for Astrophysics
Research, Science and Technology Research Institute, College Lane, AL10
9AB, Hatfield, UK }
\altaffiltext{4}{University of Turku, Tuorla Observatory, Department of 
Physics and Astronomy, V\"ais\"al\"antie 20, FI-21500, Piikki\"o, 
Finland}
\altaffiltext{5}{Department of Terrestrial Magnetism, Carnegie
Institution of Washington, 5241 Broad Branch Road, NW, Washington, DC
20015-1305, USA}
\altaffiltext{6}{Astronomy Unit, School of Mathematical Sciences, Queen 
Mary, University of London. UK}
\altaffiltext{7}{Computational Engineering and Science Research Centre, 
University of Southern Queensland, Toowoomba, Queensland 4350, 
Australia}
\altaffiltext{8}{Australian Astronomical Observatory, PO Box 915,
North Ryde, NSW 1670, Australia}
\altaffiltext{9}{The Observatories of the Carnegie Institution of 
Washington, 813 Santa Barbara Street, Pasadena, CA 91101, USA}
\altaffiltext{10}{Institute of Astrophysics, Pontificia Universidad 
Catolica de Chile, Casilla 306, Santiago 22, Chile}
\altaffiltext{11}{Vatican Observatory, V00120 Vatican City State, Italy }
\altaffiltext{12}{Departamento de Astronom{\'i}a, Universidad de Chile, 
Camino el Observatorio 1515, Las Condes, Santiago, Chile, Casilla 36-D}

\email{
rob@phys.unsw.edu.au}

\shortauthors{Wittenmyer et al.}

\begin{abstract}

\noindent We report the detection of GJ\,832c, a super-Earth orbiting 
near the inner edge of the habitable zone of GJ\,832, an M dwarf 
previously known to host a Jupiter analog in a nearly-circular 9.4-year 
orbit.  The combination of precise radial-velocity measurements from 
three telescopes reveals the presence of a planet with a period of 
35.68$\pm$0.03 days and minimum mass (m sin $i$) of 5.4$\pm$1.0 Earth 
masses.  GJ\,832c moves on a low-eccentricity orbit ($e=0.18\pm$0.13) 
towards the inner edge of the habitable zone.  However, given the large 
mass of the planet, it seems likely that it would possess a massive 
atmosphere, which may well render the planet inhospitable.  Indeed, it 
is perhaps more likely that GJ 832c is a ``super-Venus,'' featuring 
significant greenhouse forcing.  With an outer giant planet and an 
interior, potentially rocky planet, the GJ\,832 planetary system can be 
thought of as a miniature version of our own Solar system.

\end{abstract} 


\section{Introduction}

For hundreds of years, it was assumed that if planetary systems existed 
around other stars, they would look substantially like our own Solar 
system \citep{kant, laplace}.  That is, they would feature giant outer 
planets and rocky inner planets, moving on nearly-circular orbits.  The 
discovery of hundreds of extrasolar planetary systems\footnote{The 
Exoplanet Orbit Database at http://exoplanets.org} over the last 20 
years have instead revealed a picture of planetary system diversity 
``far stranger than we can imagine'' \citep{haldane}.  We now know that 
planetary systems containing a ``Jupiter analog'' (a gas giant planet which has remained in a low-eccentricity orbit beyond the ice 
line after planetary migration) are relatively uncommon 
\citep{gould10, jupiters, 2jupiters}, while results from the 
\textit{Kepler} spacecraft \citep{borucki10} have shown us that 
super-Earths in compact multiple systems are very common 
\citep{howard12, batalha13, petigura13}.  While \textit{Kepler} has 
revolutionized exoplanetary science and provided a first estimate of the 
frequency of Earth-size planets in Earth-like orbits, long-term 
radial-velocity surveys \citep{limitspaper, texas1, texas2, z13} 
complement these data with measurements of the frequency of Jupiter-like 
planets in Jupiter-like orbits.  This in turn will reveal how common 
Solar-system-like architectures are.

In addition to the finding that small, close-in planets are far more 
common than long-period gas giants \citep{howard13}, planet-search 
efforts are now expanding into new realms of parameter space, seeking to 
understand how the detailed properties of planetary systems depend on 
the properties of their host stars.  In the early days of exoplanet 
observations, host stars significantly more massive than our Sun were 
neglected, due to the difficulties in determining precise radial 
velocities.  In recent years, however, several radial velocity surveys 
have begun to take advantage of stellar evolution by observing such 
higher-mass stars once they have evolved off the main sequence to become 
subgiants and giants.  This approach has been successfully used by 
several teams \citep[e.g.][]{setiawan03, hatzes05, sato05, johnson06b, 
doellinger07, n09, 47205paper}.  

Meanwhile, at the low-mass end, M dwarfs are being targeted by a number 
of near-infrared radial-velocity surveys searching for rocky and 
potentially habitable planets \citep[e.g.][]{q10, bean10, suvrath12, 
barnes12, bonfils13}.  Notable results from these M dwarf surveys are 
that small, rocky planets are common \citep{bonfils13}, and close-in 
giant planets are rare \citep{endl06}; as yet there are no 
robust statistics on the population of longer-period giant planets.

One example of an M dwarf known to host a long-period giant 
planet is GJ\,832.  \citet{bailey09} reported the discovery by the 
Anglo-Australian Planet Search (AAPS) of a 0.64\Mjup\ planet in a 
near-circular orbit with period 9.4$\pm$0.4 yr.  The AAPS has been in 
operation for 15 years, and has achieved a long-term radial-velocity 
precision of 3 \ms\ or better since its inception, which is enabling the 
detection of long-period giant planets \citep{jones10, 142paper, 
2jupiters}.  GJ\,832b is one of only a handful of such giant planets 
known to orbit M dwarfs.  The others are GJ\,179b 
\citep{howard10}, GJ\,849b \citep{butler06,bonfils13}, GJ\,328b 
\citep{robertson13}, and OGLE-2006-BLG-109Lb \citep{gaudi08,bennett10}.  
Of the long-period giant planets known to orbit M dwarfs, GJ\,328b is 
the one with the largest separation ($a=4.5\pm$0.2 AU).  GJ\,832b, which 
lies at $a=3.4\pm$0.4 AU \citep{bailey09}, is clearly a Jupiter analog, 
and may well play a similar dynamical role in the GJ\,832 system to that 
played by Jupiter in our Solar system \citep[e.g.][]{hj08,hj09,hjc10}. 

We report here a second, super-Earth mass planet in the GJ\,832 system 
-- with a semimajor axis of $a\sim$0.16 AU, the GJ\,832 system can be 
considered a miniature Solar system analog, with an interior potentially 
rocky and habitable planet, and a distant gas giant.  This paper is 
organized as follows: Section 2 briefly describes the three data sets 
and gives the stellar parameters.  Section 3 details the traditional and 
Bayesian orbit fitting procedures, and gives the parameters of GJ\,832c.  
In Section 4, we give a discussion on potential habitability before 
drawing our final conclusions.

\section{Observational Data}

We have combined three high-precision radial-velocity data sets that 
span a variety of baselines.  The data covering the longest baseline (39 
epochs over 15 years) were taken by the Anglo-Australian Planet Search 
(AAPS) team, using the UCLES echelle spectrograph \citep{diego:90}.  An 
iodine absorption cell provides wavelength calibration from 5000 to 
6200\,\AA.  The spectrograph point-spread function (PSF) and wavelength 
calibration are derived from the absorption lines embedded on the 
spectrum by the iodine cell \citep{val:95,BuMaWi96}.  The result is a 
precise Doppler velocity estimate for each epoch, along with an internal 
uncertainty estimate, which includes the effects of photon-counting 
uncertainties, residual errors in the spectrograph PSF model, and 
variation in the underlying spectrum between the iodine-free template 
and epoch spectra observed through the iodine cell.  All velocities are 
measured relative to the zero-point defined by the template observation.  
GJ\,832 has been observed on 39 epochs since (Table~\ref{AATvels}), with 
a total data span of 5465 d (15 yr).  The mean internal velocity 
uncertainty for these data is 2.6\,\ms.

GJ\,832 has also been observed with the Planet Finder Spectrograph (PFS) 
\citep{crane06,crane08,crane10} on the 6.5m Magellan II (Clay) 
telescope.  The PFS is a high-resolution ($R\sim\,80,000$) echelle 
spectrograph optimised for high-precision radial-velocity measurements 
\citep[e.g.][]{albrecht11, albrecht12, gj667, arr13}.  The PFS also uses 
the iodine cell method as descibed above.  The 16 measurements of 
GJ\,832 are given in Table~\ref{PFSvels}.  The data span 818 days and 
have a mean internal uncertainty of 0.9\,\ms.  A further 54 velocities 
were obtained from a HARPS-TERRA \citep{ang12b} reduction of the 
publicly available spectra and and are given in Table~\ref{HARPSvels}.  
The AAPS data are critical for constraining the long-period outer 
planet, and the extremely precise HARPS and PFS data are necessary for 
characterizing the inner planet.

To account for possible intrinsic correlations in the radial velocities, 
we used the same statistical model as \citet{tuomi2014} that assumes 
that the deviation of the $i$th measurement of a given instrument from 
the mean depends also on the deviation of the previous measurement.  In 
other words, these deviations are correlated with a correlation 
coefficient of $\phi \exp \big\{ -(t_{i} - t_{i-1})/\tau \big\}$ that 
decreases exponentially as the gap between the two measurements (the 
difference $t_{i} - t_{i-1}$) increases.  We set the correlation 
time-scale such that $\tau = 4$ days \citep{baluev2013,tuomi2014b}. 
Parameter $\phi \in [-1,1]$ is a free parameter in the model of 
\citet{tuomi2014b}.  The maximum \textit{a posteriori} estimates of 
these ``nuisance parameters'' are given in 
Table~\ref{instrument_parameters}.  Only for the HARPS-TERRA data does 
the correlation coefficient $\phi$ differ significantly from zero.  We 
thus consider an additional data set in our analysis, ``HARPS-CR,'' 
which has been corrected for these intrinsic correlations.  Those data 
are also shown in Table~\ref{HARPSvels}.

\section{Orbit Fitting and Planetary Parameters}

GJ\,832 (HD\,204961, HIP\,106440, LHS\,3685) is a very nearby M1.5 
dwarf, lying at 4.95 pc \citep{vl07}.  The parameters of the host star 
are summarized in Table~\ref{stellarparams}.  While the original 
discovery paper for the giant planet \citep{bailey09} did not note any 
residual signals of interest, \citet{bonfils13} combined AAT ($N=32$) 
and HARPS ($N=54$) data to refine the planet's orbital parameters and 
mentioned a potential 35-day residual periodicity.  They concluded that 
the data in hand did not yet warrant a secure detection as the 
false-alarm probability (FAP) exceeded 1\%.  The AAT data published in 
\citet{bailey09} contained 32 epochs spanning 3519 d (9.6 yr).  
Combining all available data, we now have 109 epochs covering a 15-year 
baseline, enabling us to better characterize the long-period planet, and 
increasing our sensitivity to any residual signals of interest.

\subsection{Bayesian approach}

We analysed the combined HARPS, PFS, and UCLES radial velocities in 
several stages.  First, we drew a sample from the posterior density of a 
model without Keplerian signals, i.e.~a model with $k=0$, by using the 
adaptive Metropolis algorithm \citep{haario2001}.  This is a 
generalization of the Metropolis-Hastings Markov chain Monte Carlo 
technique \citep{metropolis1953,hastings1970} that adapts the proposal 
density to the information gathered from the posterior.  This baseline 
model enabled us to determine whether the models with $k=1, 2, ...n$ 
Keplerian signals were, statistically, significantly better descriptions 
of the data by estimating the Bayesian evidence ratios of the different 
models \citep[e.g.][]{kass1995,tuomi2014,tuomi2014b}. For this purpose, 
we used the simple estimate described in \citet{newton1994}.

The search for periodic signals in the data was performed by using 
tempered samplings \citep{tuomi2013c,tuomi2014,tuomi2014b} such that a 
scaled likelihood $l(m | \theta)^{\beta}$ and a scaled prior density 
$\pi(\theta)^{\beta}$, where $m$ is the measurements and $\theta$ the 
parameter vector, instead of the common likelihood $l(m | \theta)$ and 
prior $\pi{\theta}$.  We choose $\beta \in (0,1)$ low enough such that 
the posterior probability density is scaled sufficiently to enable the 
Markov chains to visit repeatedly all relevant areas in the parameter 
space, the period space in particular.  In this way, we can estimate the 
general shape of the posterior density as a function of the period 
parameter to see which periods correspond to the highest probability 
maxima.  This is not necessarily possible with ``normal'' samplings (i.e. 
when $\beta = 1$) because one or some of the maxima in the period space 
could be so high and significant that the Markov chains fail to visit the 
whole period space efficiently due to the samplings getting ``stuck'' in 
one of the corresponding maxima. This could happen because the parameter 
space around some maximum is in practice so much less probable that any 
proposed values outside the maximum are rejected in the MCMC sampling. 
The application of such tempered samplings thus enables an efficient 
search for periodicities in the data.

\begin{figure}
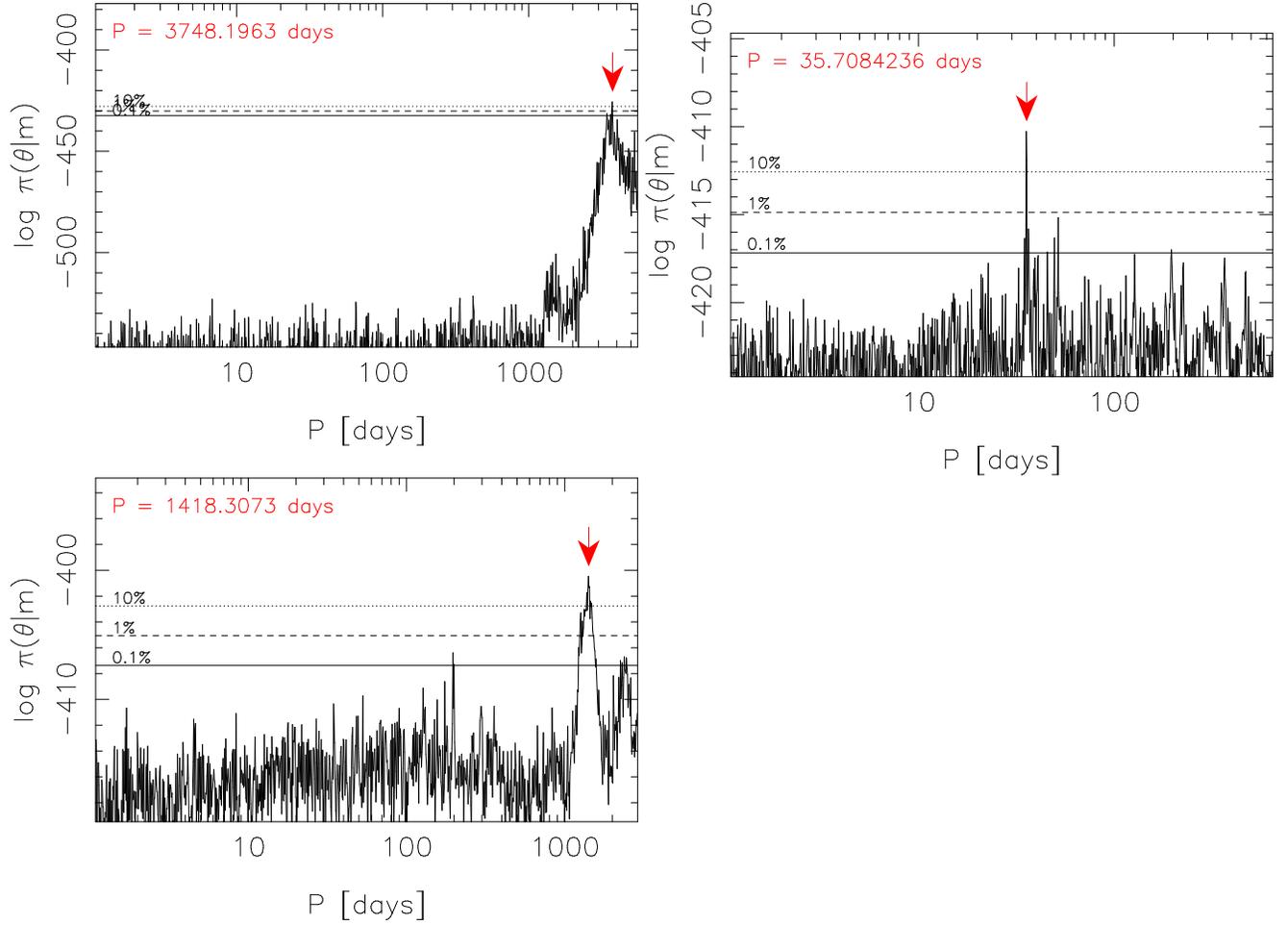

\includegraphics[angle=270.0,scale=.40]{rv_GJ832_01_pcurve_b.ps}
\includegraphics[angle=270.0,scale=.40]{rvdist02_GJ832_pcurve_c.ps}
\includegraphics[angle=270.0,scale=.40]{rv_GJ832_03_pcurve_d.ps}
\caption{Estimated posterior density of the period of the Keplerian 
signals based on MCMC sampling. The red arrow indicates the global 
maximum identified by the chain and the horizontal lines denote the 10\% 
(dotted), 1\% (dashed), and 0.1\% (solid) probability thresholds with 
respect to the maximum.  Top panel: GJ\,832b. Middle panel: GJ\,832c. 
Bottom panel: Residuals to two-planet fit; this periodicity did not meet 
our criteria for a significant detection. }
\label{dram}
\end{figure}

We maximised parameter $\beta \in (0,1)$ such that the period parameter 
visited all areas in the period space between one day and the data 
baseline during these tempered samplings.  The results from this period 
search reveal the shape of the posterior probability density as a 
function of period (see Figure~\ref{dram}).  This enabled us to identify 
all relevant maxima in the period space because such transformation 
artificially decreases the significances of the maxima, while leaving 
their locations unchanged.  To ensure that the Markov chains visited all 
areas of high posterior probability in the parameter space, and 
especially through the period space, we applied the delayed-rejection 
adaptive-Metropolis algorithm \citep{haario06}, where another proposed 
parameter vector from a narrower proposal density is tested if the first 
proposed vector is rejected.  This enables an efficient periodicity 
search because the chains visit the narrow probability maxima in the 
period space.  One such sampling is shown in the middle panel of Figure~\ref{dram} when 
searching for a second periodicity in the data.  The chain clearly 
identifies a global maximum at a period of 35.7 days.

\begin{figure}
\includegraphics[angle=270.0,scale=.60]{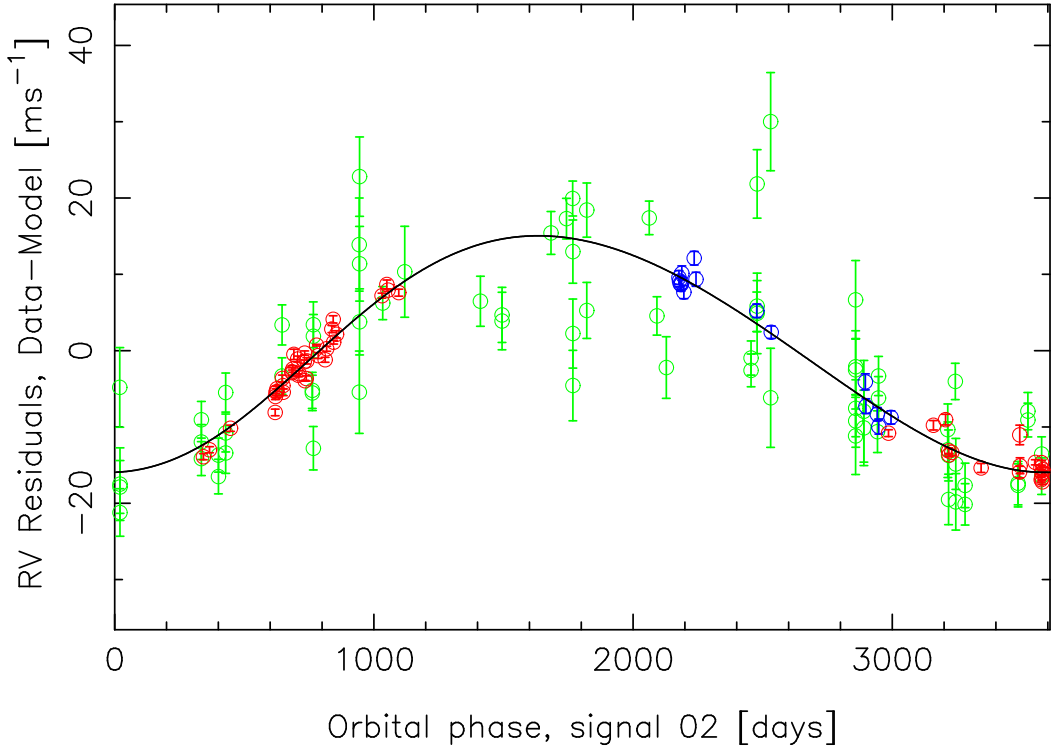}
\caption{Radial velocities and fit for GJ\,832b; the signal of the 
second planet has been removed.  AAT -- green, HARPS -- red, PFS -- 
blue. }
\label{mikkorvplots1}
\end{figure}

\begin{figure}
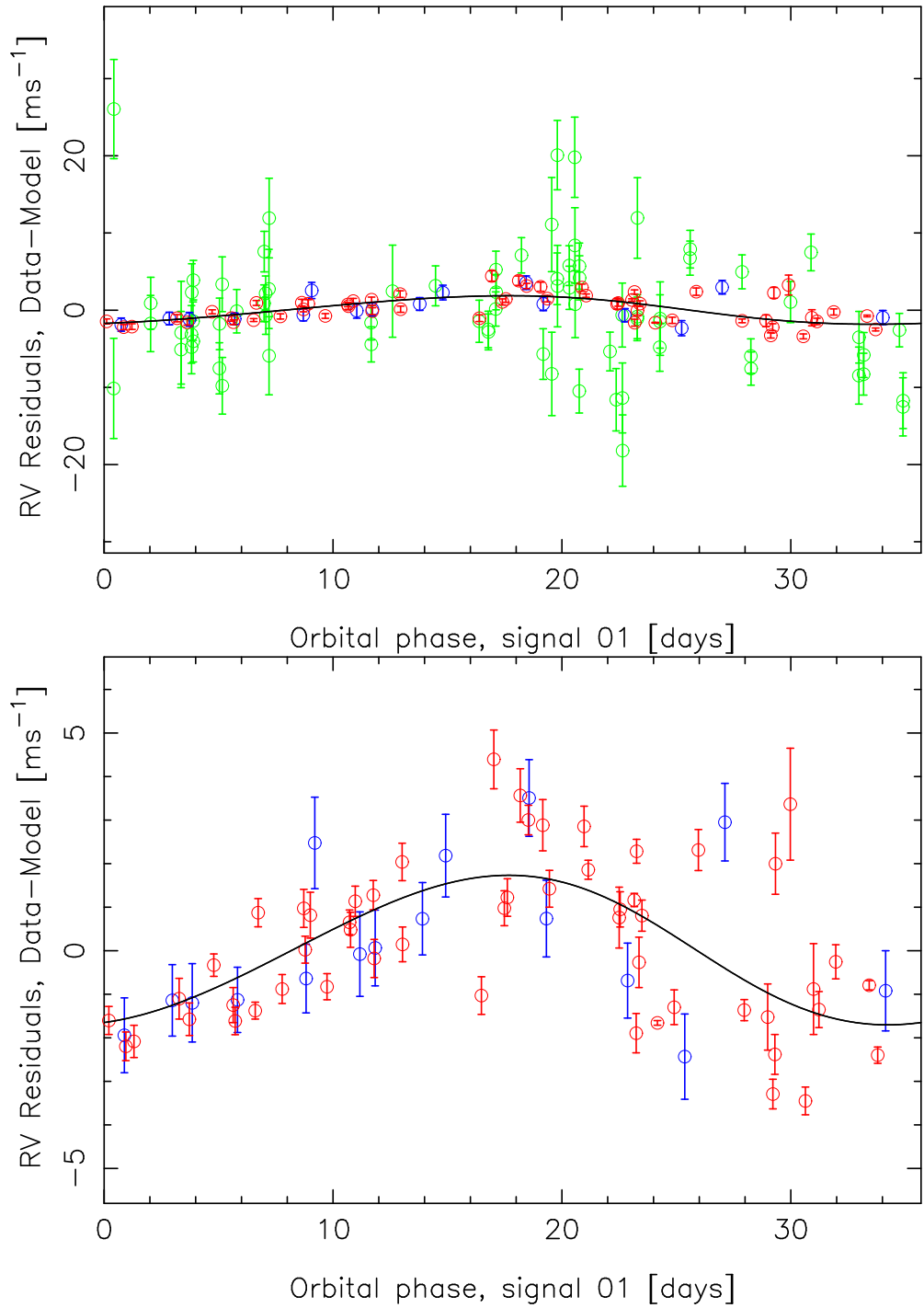

\includegraphics[angle=270.0,scale=.57]{rvdist02_scresidc_GJ832_1.ps}
\includegraphics[angle=270.0,scale=.57]{rv_NO_AATdist02_scresidc_GJ832_1.ps}
\caption{Top: Radial velocities and fit for GJ\,832c; the signal of the 
outer planet has been removed.  AAT -- green, HARPS -- red, PFS -- blue.  
Bottom: Same, but the AAT data have been omitted from the plot to more 
clearly show the low-amplitude signal. }
\label{mikkorvplots2}
\end{figure}

After such tempered samplings, we started several ``cold chains,'' 
i.e.~normal chains such that $\beta=1$, in the vicinity of the highest 
maxima to determine which of them were significant according to the 
detection criteria discussed in \citet{tuomi2014} and 
\citet{tuomi2014b}.  The statistical significance of a signal is 
quantified by the Bayesian evidence ratio $B(k,k-1)$.  We required that 
the Bayesian evidence ratio be at least $10^{4}$ times greater for a 
model with $k+1$ than for a model with $k$ signals to state that there 
are $k+1$ signals present in the data.  That is, the model with 
the signal must be 10000 times more probable than the model without.  
For the combination of the three data sets considered here, we obtain 
$B(1,0)=4.5\times\,10^{60}$ (in favor of a one-planet model over zero 
planets).  The 35-day signal is also significant with respect to our 
detection threshold as it is detected with $B(2,1)=6.6\times\,10^5$.  
The maximum \textit{a posteriori} estimates of the model parameters, 
together with the corresponding Bayesian 99\% credibility intervals, are 
listed in Table~\ref{planetary_parameters}. The data and best-fit models 
are shown in Figure~\ref{mikkorvplots1} (GJ\,832b) and 
Figure~\ref{mikkorvplots2} (GJ\,832c).

In addition to the velocities, various activity indices are also 
available for the epochs of the GJ\,832 HARPS observations 
\citep{bonfils13}.  Those metrics, derived from the cross-correlation 
function (CCF) are the bisector inverse slope (BIS), described fully in 
\citet{queloz01}, and the CCF full width at half maximum (FWHM).  We 
modelled correlations of the HARPS-TERRA velocities with BIS, FWHM, and 
$S$-index by assuming these correlations were linear that represents the 
first-order approximation for such dependence.  However, accounting for 
these correlations did not improve the model, which indicates that such 
linear correlations were insignificant.  Furthermore, when we 
computed the best-fit estimates for such linear correlation parameters, 
they were all consistent with zero, as shown in Table~\ref{activity}.

\subsection{Traditional approach}

The combination of three data sets with high precision and long 
observational baseline yields evidence for a second, low-mass planet 
orbiting GJ\,832.  Given that we have used data from every telescope 
which is able to achieve sufficient velocity precision for this 
Southern M dwarf ($\delta=-49.0$\degrees), independent confirmation of 
GJ\,832c is problematic.  It is prudent, then, to employ an independent 
analysis to test the plausibility of the 35-day signal.

For this analysis, we use the HARPS data set which has been corrected 
for intrinsic correlations as described above -- labeled here as 
``HARPS-CR.'' An instrumental noise term was derived from the excess 
white noise parameter given in Table~\ref{instrument_parameters}-- 
HARPS-CR: 1.33\ms, PFS: 1.45\ms, AAT: 4.66\ms.  Before orbit fitting, 
that noise was added in quadrature to the uncertainties of each data 
point.  We repeated our analysis using the unaltered HARPS-TERRA 
velocities, and found throughout that the HARPS-CR and HARPS-TERRA data 
sets gave the same results.

\begin{figure}
\plotone{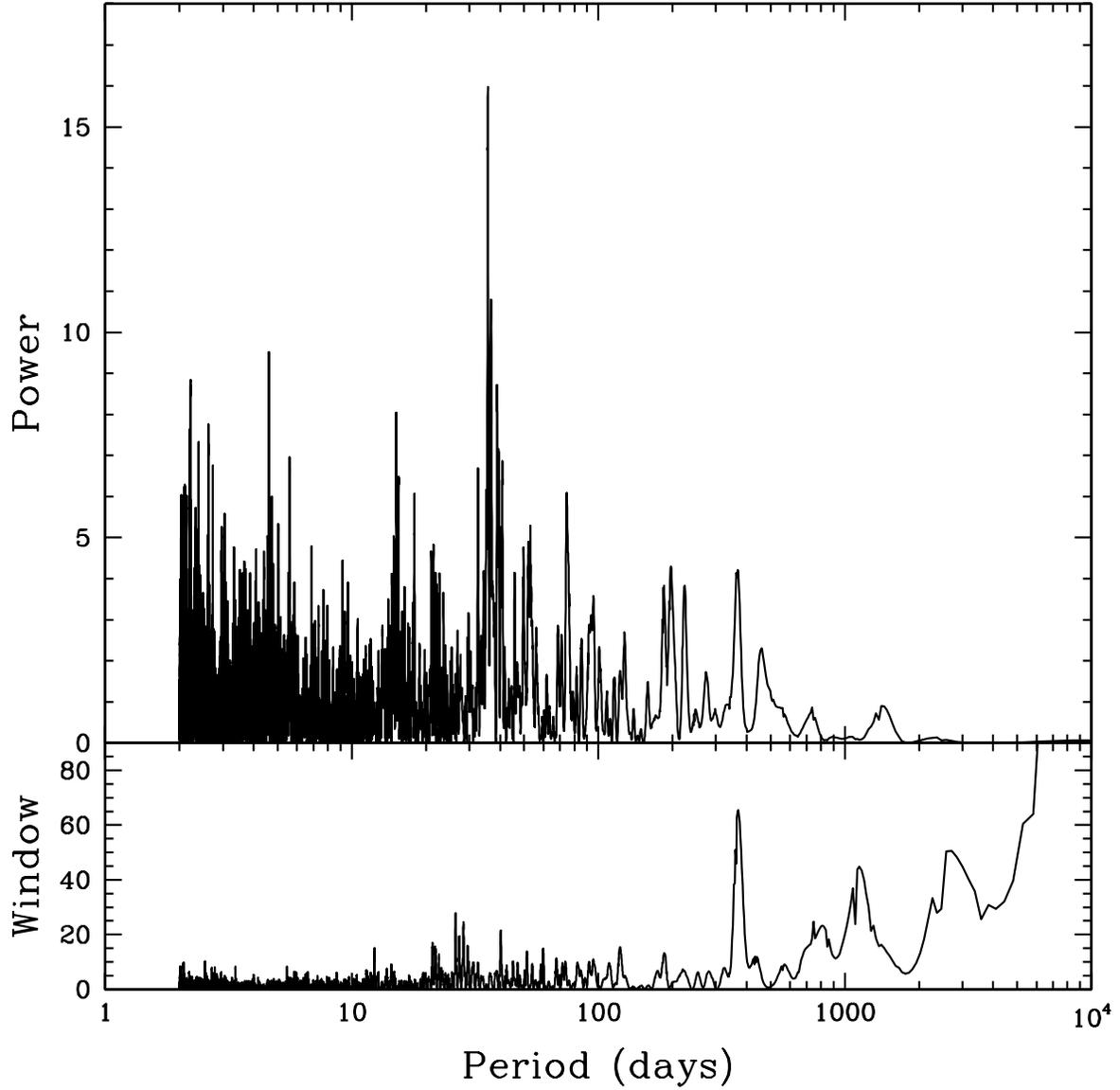}
\caption{Generalized Lomb-Scargle periodogram of the residuals to a 
single-planet fit for GJ\,832.  A strong peak at 35.67 days is present, 
with a bootstrap FAP of 0.04\%. }
\label{1planetpgram}
\end{figure}

First, we fit a single-planet model to the three data sets using the 
nonlinear least-squares minimization routine \textit{GaussFit} 
\citep{jefferys88}.  The velocity offsets between the three data sets 
were included as free parameters.  The rms scatter about the three data 
sets are as follows -- AAT: 5.72\,\ms, HARPS-CR: 1.88\,\ms, PFS: 
1.74\,\ms.  We performed a periodogram search on the residuals to the 
one-planet fit, using the generalized Lomb-Scargle formalism of 
\citet{zk09}. Unlike the classical Lomb-Scargle periodogram 
\citep{lomb76, scargle82}, this technique accounts for the uncertainties 
on the individual data points, which is critically important for the 
case of GJ\,832 where we have combined data sets with significantly 
different precisions.  The periodogram of the 1-planet fit is shown in 
Figure~\ref{1planetpgram}; the highest peak is at 35.67 days.  The FAP 
was estimated using a bootstrap randomization method \citep{kurster97}.  
From 10,000 bootstrap realizations, the 35.67-day peak is shown to be 
highly significant, with FAP=0.0004 (0.04\%).


We then used a genetic algorithm to search a wide parameter space for 
two-planet models, and to check that any candidate secondary signal is 
indeed the global best-fit.  Our group has used this approach 
extensively \citep[e.g.][]{tinney11,HUAqr,hkpaper} when the orbital 
parameters of a planet candidate are highly uncertain.  We allowed the 
genetic algorithm to fit 2-Keplerian models to the three data sets 
simultaneously, searching secondary periods from 10 to 3000 days.  It 
ran for 50,000 iterations, testing a total of about $10^7$ possible 
2-Keplerian configurations.  The genetic algorithm converged on 
$P_2\sim$35 days, giving confidence that this is the most likely period 
for a candidate second planet.

\begin{figure}
\plotone{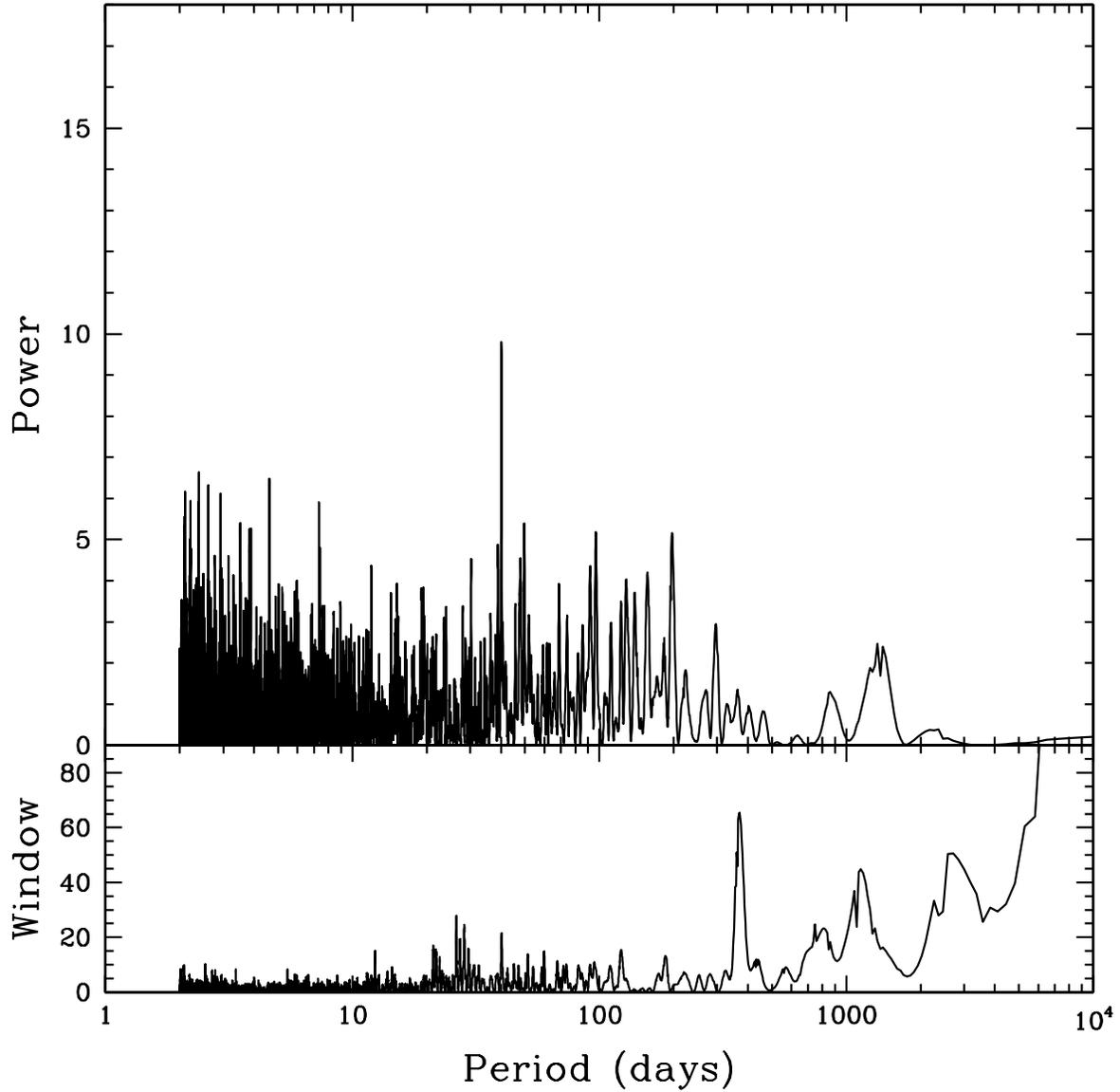}
\caption{Generalized Lomb-Scargle periodogram of the residuals to the 
two-planet fit for GJ\,832.  All three datasets are included.  The peak 
at 40 days has a bootstrap FAP of 4.56\%, 100 times less significant 
than the peak due to the inner planet (Figure~\ref{1planetpgram}). }
\label{2plresids}
\end{figure}

We then obtained a final 2-planet fit using \textit{GaussFit}.  Again, 
we performed the fit twice, using the two versions of the HARPS 
velocities.  The details of each fit are summarized in Table~\ref{rms}. 
Both fits gave the same results, though the HARPS-CR set (``Fit 2'') 
gave a slightly better rms and smaller uncertainties on the planetary 
parameters - hence, we adopt those results in Table~\ref{planetparams}.  
These fits reveal a second planet, GJ\,832c, with $P=35.68\pm$0.03 d and 
m~sin~$i=5.40\pm$0.95\Mearth\ on a nearly-circular orbit.  A periodogram 
of the residuals to the 2-planet fit is shown in Figure~\ref{2plresids}; 
the highest peak at 40.2 days has a bootstrap FAP of 0.0456 (4.6\%).

\section{Discussion and Conclusions}

\subsection{Testing the planet hypothesis}

If the 35-day signal is real, adding data should result in a higher 
significance level, i.e. a lower FAP determined by the bootstrap method 
described above \citep{kurster97}.  We test this by performing 
one-planet fits on various combinations of the three data sets 
considered here.  In these fits, the parameters of the outer 
planet are started at the best-fit values in Table~\ref{faps}, but are 
allowed to vary. For data combinations with insufficient time baseline 
to adequately fit the outer planet, its parameters are instead fixed at 
their best-fit values.  After each fit, we removed the signal of the 
outer planet, examined the periodogram \citep{zk09} of the residuals, 
and computed the FAP of the highest remaining peak using 10,000 
bootstrap randomizations.  The results are summarized in 
Table~\ref{faps}.  The AAT data alone are not sufficiently precise to 
detect the $K\ltsimeq$2\ms\ signal of the candidate planet, nor did the 
addition of only 16 epochs from PFS enable the detection of any 
significant residual signals.  Nevertheless, we see in Table~\ref{faps} 
that the addition of data indeed strengthens the significance of the 
35.6-day signal, adding confidence that the signal is real and not an 
artifact of one particular instrument.  Table~\ref{faps} also 
indicates that the HARPS data are necessary to pull out the signal of 
GJ\,832c, and the AAT data, while noisy, are necessary for constraining 
the outer planet. 

\begin{figure}
\plottwo{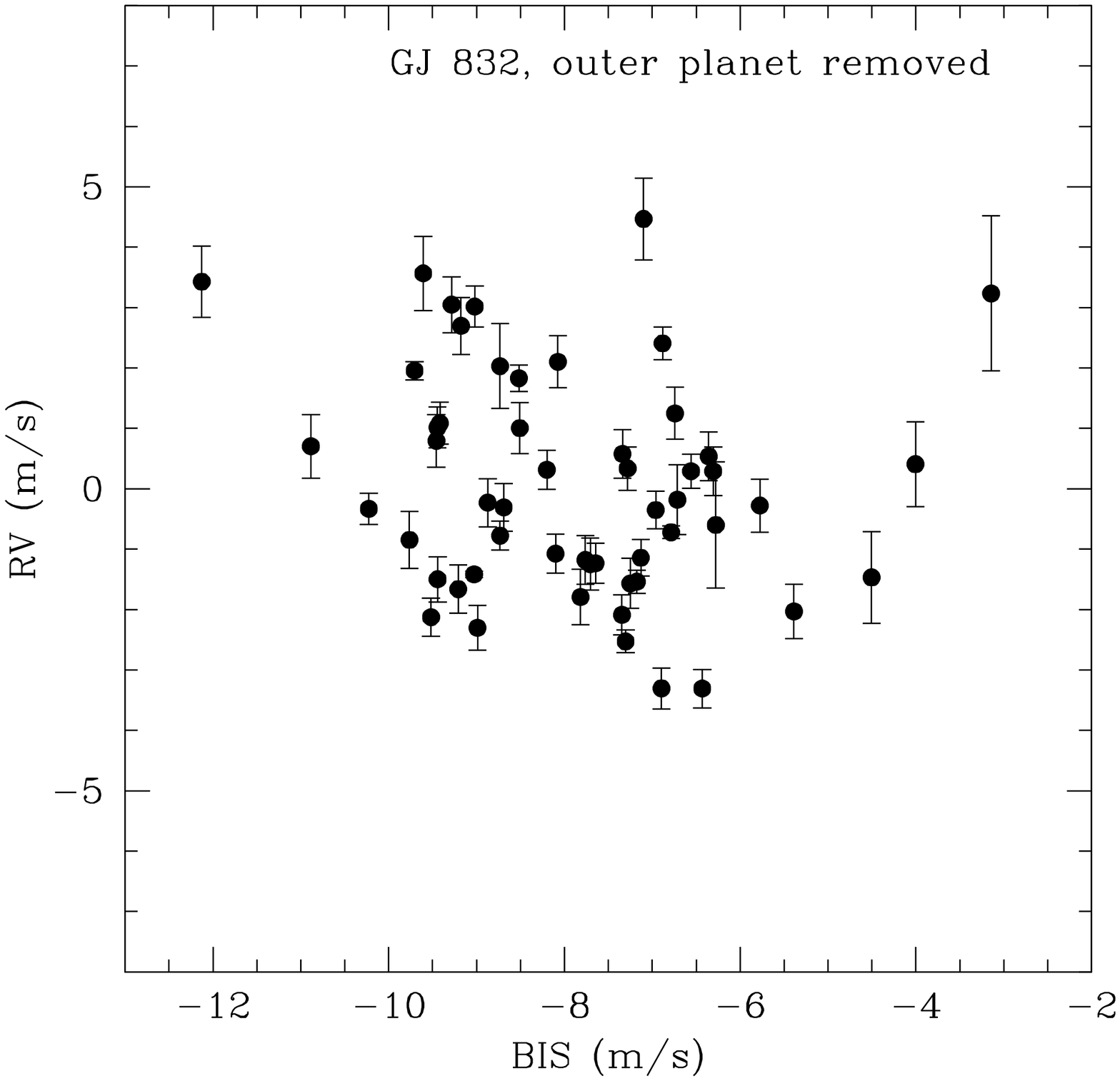}{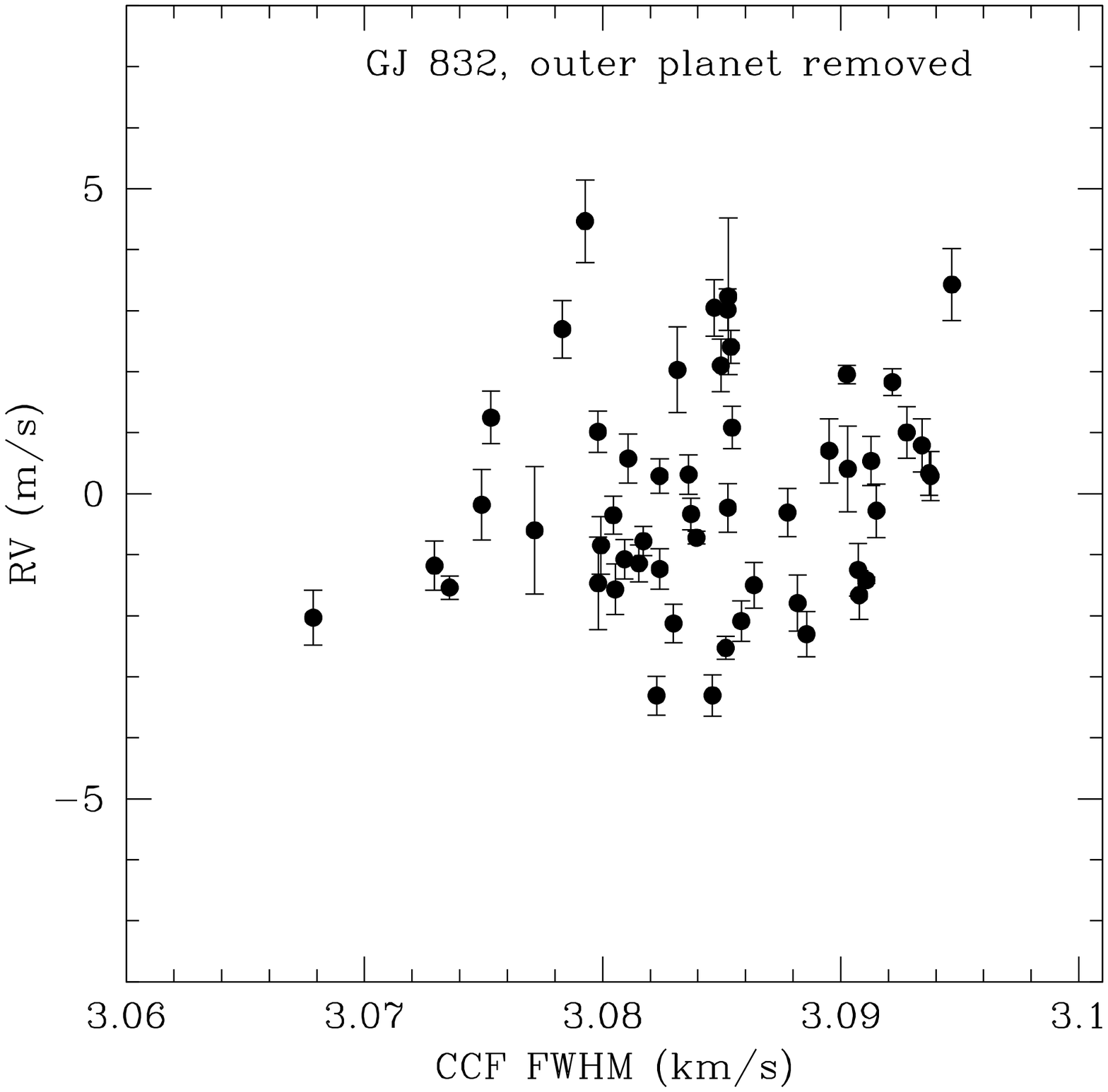}
\caption{HARPS radial velocities (after removing the signal of GJ\,832b) 
versus the bisector inverse slope (BIS: left panel) and the FWHM of the 
cross-correlation function (CCF FWHM: right panel).  No correlations are 
evident, supporting the hypothesis that the 35.6-day signal is due to an 
orbiting planet. }
\label{bis}
\end{figure}

The next obvious question to ask is whether the detected signal is 
intrinsic to the star.  As noted in \citet{bonfils13} and in Section 
3.1, the HARPS planet-search programs use the additional diagnostics BIS 
and CCF-FWHM to check for star-induced variability.  Being 
contemporaneous with the velocity measurements, both of these measures 
can be directly compared with the velocity derived from a given 
spectrum.  If BIS or FWHM show correlations with the velocities, a 
candidate radial-velocity signal can be considered suspect.  
Figure~\ref{bis} plots the HARPS velocities (after removing the outer 
planet) against the BIS (left panel) and the CCF FWHM (right panel).  No 
correlations are evident, and the highest BIS periodogram peak at 179.6 
days has a bootstrap FAP of 17\%.  For FWHM, the highest periodogram 
peak at 6322 days has a bootstrap FAP$<$0.01\%.  For comparison, 
the outer planet has a period of 3660 days, and the HARPS FWHM data only 
span 1719 days; for these reasons, we maintain the conclusion of 
\citet{bailey09} that the long-period velocity signal is due to an 
orbiting body.  As shown in Table~\ref{activity}, none of 
these activity indicators had correlations significantly different from 
zero.  These results are further evidence that the 35.6-day signal is 
not intrinsic to the star.

\subsection{GJ\,832c: a habitable-zone super-Earth}

In recent years, a growing number of super-Earths have been discovered 
that orbit their host stars at a distance that may be compatible with 
the existence of liquid water somewhere on the planet were it to have a 
surface (i.e.~within the classical habitable zone).  A list of these 
planets is given in Table~\ref{habitable}.  Of those planets, perhaps 
the most interesting are those orbiting GJ\,581.  In that system, a 
total of six planets have been claimed, although at least two of these 
are still the subject of significant debate (e.g. Mayor et al. 2009, 
Vogt et al. 2010, Tuomi 2011, von Braun et al. 2011, Tadeu dos Santos et 
al. 2012, Vogt et al. 2012, Baluev 2013).  The proposed planetary system 
around GJ\,581 displays an orbital architecture that is strikingly 
similar to a miniature version of our own Solar system.  The similarity 
to our own Solar system has recently been enhanced by the results of the 
DEBRIS survey (Matthews et al. 2010), which recently discovered and 
spatially resolved a disk of debris orbiting GJ\,581 (Lestrade et al. 
2012), analogous to the Solar system's Edgeworth-Kuiper belt.

We can estimate the location of the classical habitable zone following 
the prescriptions given in Selsis et al. (2007) and Kopparapu et al. 
(2014), using the stellar parameters detailed in 
Table~\ref{stellarparams}. In both cases, we find that GJ\,832c lies 
just inside the inner edge of the potentially habitable region - with 
the Selsis et al. prescription yielding a habitable zone that stretches 
between 0.13 and 0.28 au, and the Kopparapu et al. prescription 
suggesting that the conservative habitable zone for a 5\Mearth\ 
planet lies between 0.130 and 0.237 au, compared to the measured 
$a=0.163\pm$0.006 au for GJ\,832c.

Although GJ\,832c is sufficiently far from its host star that there is 
the potential for liquid water to exist on its surface, this does not 
necessarily make that planet truly habitable.  Indeed, there is a vast 
number of factors that can contribute to the habitability of a given 
exoplanet beyond the distance at which it orbits its host star (e.g. 
Horner \& Jones 2010, Horner 2014).  Given the planet's proximity to its 
host star, it seems likely that GJ\,832 c will be trapped in a 
spin-orbit resonance, though moderate orbital eccentricity may mean it 
is not necessarily trapped in a resonance that causes one side of the 
planet to perpetually face towards the Sun.  In our own Solar system, 
the planet Mercury is trapped in such a resonance: rotating three times 
on its axis in the time it takes to complete two full orbits of the Sun.  
Mercury's capture to that particular resonance is almost certainly the 
result of its relatively eccentric orbit (e.g. Correia \& Laskar 2009) - 
and so it is certainly feasible that GJ\,832 c, whilst tidally locked, 
is not trapped in 1:1 spin-orbit resonance.  Even if the planet is 
trapped in such a resonance, however, that might not be deleterious to 
the prospects for its being habitable.  For example, recent work 
by \citet{yang14}, employing a 3-dimensional general circulation model, 
suggests that planets with slower rotation rates would be able to remain 
habitable at higher flux levels than for comparable, rapidly rotating 
planets. 

Given the planet's large mass, however, it is likely to be shrouded in a 
dense atmosphere - which might in turn render it uninhabitable.  In that 
scenario, the dense atmosphere would provide a strong greenhouse effect, 
raising the surface temperature enough to cause any oceans to boil away, 
as is thought to have happened to Venus early in the lifetime of the 
Solar system, e.g. Kasting (1988).  Kasting et al. (1993) proposed that 
tidally locked planets around late-type stars might be rendered 
uninhabitable by atmospheric freeze-out if they were locked in a 1:1 
spin-orbit resonance.  However, such a massive atmosphere would also be 
able to prevent the freeze out of the planet's atmosphere if it were 
trapped in a 1:1 spin-orbit resonance (Heath et al. 1999).  A detailed 
review of the potential habitability of planets around M dwarfs by 
Tarter et al. (2007) re-opened the possibility of habitability for such 
planets.  More recently, \citet{kop14} have argued that the 
inner edge of the habitable zone moves inward for more massive planets 
-- a scenario which would operate in favor of GJ\,832c's habitability.

Given the large mass of GJ\,832c, and the high probability of it having 
a thick, dense atmosphere, it is reasonable to assume that it is 
unlikely to be a habitable planet.  However, it is natural to ask 
whether it could host a giant satellite, which might itself be 
habitable.  Speculation about habitable exomoons is not a new thing, and 
in recent years, a number of papers have been published discussing the 
prospects for such satellites orbiting a variety of newly discovered 
planets -- e.g. the gas giants HD\,38283b (Tinney et al. 2011), and 
HD\,23079b (Cuntz et al. 2013), and the super-Earth Kepler-22b (Kipping 
et al. 2013), or discussing the viability of such satellites as 
potential locations for life in a more general sense (e.g. Heller 2012, 
Forgan \& Kipping 2013, Heller \& Barnes 2013).  As such, it is 
interesting to consider whether GJ\,832c could host such a 
potentially-habitable satellite, although we acknowledge that the 
detection of such a satellite is currently well beyond our means.

Within our Solar system, one planet (the Earth) and several of the minor 
bodies (such as Pluto) are known to have giant satellites that are 
thought to have formed as a result of giant impacts on their host object 
toward the end of planet formation (e.g. Benz et al. 1986, Canup 2004, 
Canup 2005).  In the case of the Pluto-Charon binary, the mass of Charon 
is approximately 1/9th that of Pluto - were that extrapolated to the 
case of GJ\,832c, it would result in a moon somewhat greater than half 
of the mass of the Earth\footnote{We note that the m sin\,$i$ 
determined from radial velocity measurements would actually be the total 
mass of the exoplanet in question, plus any moons it hosts.  For 
example, if GJ\,832c hosts a moon with 20\% of its own mass, then the 
planet mass would actually be only 0.8 times the m sin\,$i$ given in 
Table~\ref{planetparams}.}.  But could GJ\,832c retain such a satellite 
whilst orbiting so close to its host star?

The Hill sphere of an object is the region in which its gravitational 
pull on a satellite (or passing object) would dominate over that from 
any other object.  Typically, within our Solar system, the regular 
satellites of the planets orbit their hosts well within their Hill 
sphere.  Our Moon, for example, orbits at approximately one-quarter of 
the Hill radius.  For GJ\,832c, assuming a mass of 5.406 times that of 
the Earth, and a host-star mass of 0.45\Msun, the Hill radius would be 
just 0.00306 au - or 460,000 km.  In and of itself, this result does not 
seem to preclude the existence of a habitable exomoon orbiting GJ\,832c.  
However, Cuntz et al.(2013) found that, for the case of the gas giant 
planet HD\,23079b, satellites on prograde orbits were only stable out to 
a distance of approximately 0.3 Hill radii - a result that compares 
relatively well to the orbital distance of the Moon, which currently 
orbits Earth at a distance of $\sim0.25$ Hill radii.  Were the same true 
for the case of GJ\,832c, this would reduce the region of stability, 
requiring that a satellite orbiting that planet must remain within an 
orbital radius of $\sim138,000$ km in order to remain bound on 
astronomically long timescales.

Heller \& Barnes (2013) consider the possibility of habitable exomoons 
orbiting the super-Earth Kepler-22b and the gas giant planet candidate 
KOI211.01.  By considering the influence of tidal heating on the 
potential satellites of these planets, they reach the conclusion that 
``If either planet hosted a satellite at a distance greater than 10 
planetary radii, then this could indicate the presence of a habitable 
moon.'' If we assume that GJ\,832c is a predominantly rocky/metallic 
object, then we can obtain a rough estimate of its radius by following 
Seager et al. (2007).  For a silicaceous composition, given a mass of 
approximately 5\Mearth, it seems likely that GJ\,832c would have a 
radius approximately fifty percent greater than that of the Earth, or 
approximately 10,000 km.  We can therefore determine a rough inner-edge 
to the circum-planetary habitable zone for GJ\,832c, at approximately 10 
times this value.  In other words, for GJ\,832c to host a habitable 
exomoon, potentially formed by means of a giant collision during the 
latter stages of planet formation, such a satellite would most likely 
have to orbit between ~$\sim100,000$ and $\sim138,000$ km - a very 
narrow range.  Although the idea of a habitable exomoon companion to 
GJ\,832c is certainly interesting, the odds seem stacked against the 
existence of such an object.

\subsection{Conclusions}

We have combined high-precision radial-velocity data from three 
telescopes to detect a super-Earth (5.4$\pm$1.0 \Mearth) orbiting 
GJ\,832 near the inner edge of the habitable zone.  We attribute 
this detection to two key differences from the \citet{bonfils13} 
analysis.  The first is that our Bayesian techniques are better at 
picking out weak signals; this was powerfully demonstrated by 
\citet{tuomi2013c}, who used this approach for the GJ\,163 system and 
obtained results consistent with \citet{bonfils13} with only 35\% of the 
HARPS data used in the discovery work.  The second is that the 
HARPS-TERRA velocities are more sensitive to planet c than the 
velocities derived by the HARPS team in \citet{bonfils13}; 
\citet{ang12b} showed that HARPS-TERRA produces better velocities for M 
dwarfs. 

 Given GJ\,832's close proximity it is bright enough for high contrast 
imaging \citep{salter14}, even though it is an M-dwarf.  However, due to 
its likely old, though uncertain, age even GJ\,832b (0.63\Mjup) would 
not be bright enough to be detected by the current state of the art 
instruments such as the Gemini Planet Imager on Gemini South 
\citep{mac08}.  With a rare Jupiter analog and a potentially rocky inner 
planet, the GJ\,832 system can be considered a scaled-down version of 
our Solar system.  With this in mind, it would be interesting to see if 
that analogy continues beyond the planetary members of the system to the 
debris.  There is a growing body of work, based on \textit{Spitzer} and 
\textit{Herschel} observations, revealing correlations between the 
presence of debris disks and planets \citep[][e.g. 
]{wyatt12,maldonado12,bryden13}.  As GJ\,832 is very nearby (4.95 pc), 
it is an ideal candidate for future imaging efforts to search for debris 
disks akin to our own Edgeworth-Kuiper belt and main asteroid belt.  
Recent work by \citet{marshall14} showed a correlation between 
low (sub-Solar) metallicity, low mass planets and an elevated incidence 
of debris from \textit{Herchel} data and radial-velocity results.  As 
GJ\,832 is quite metal-poor ($[Fe/H]=-0.3$), it would thus appear to be 
a promising target for debris detection.  Future observations of 
GJ\,832 hold the promise to yield up further secrets from this 
intriguing system.

Circumstellar debris discs around mature stars are the byproduct of a 
planetesimal formation process, composed of icy and rocky bodies ranging 
from micron sized grains to kilometre sized asteroids (see reviews by 
e.g. Wyatt 2008, Krivov 2010 and Moro-Martin 2013).  The dust we 
actually observe is continually replenished in the disc through the 
collisional grinding of planetesimals as the grains are much shorter 
lived than the age of the host star, being removed by radiative 
processes and collisional destruction (Backman \& Paresce 1993).  Since 
planets are believed to be produced through the hierarchical growth of 
planetesimals from smaller bodies and dust grains are produced through 
their collisional destruction, we expect the two phenomena to be linked.  
The solar system represents one outcome of the planet formation process, 
comprising four telluric planets, four giant planets and two debris 
belts - the inner, warm Asteroid belt at 3 AU (Backman et al. 1995) and 
the outer, cold Edgeworth-Kuiper belt at 30 AU (Vitense et al. 2012).  
We detect thermal emission from warm dust around 2$\pm$2\% of sun-like 
stars (Trilling et al. 2008) and cold dust around 20$\pm$2\% of sun-like 
stars (Eiroa et al. 2013); such measurements are limited by sensitivity 
particularly for the warm dust.  By comparison, the solar system's 
debris disc is atypically faint, expected to lie in the bottom few 
percent of disc systems (Greaves \& Wyatt 2010) and currently beyond the 
reach of direct detection by ground or satellite observatories.  Around 
other stars, many host infrared excesses with two characteristic 
temperatures, typically at $\sim$150 K and $\sim$50 K (Morales et al. 
2011).  Drawing an analogy to the solar system, discs with two 
temperature components are interpreted as being the product of 
physically distinct debris belts at different orbital radii.  Due to the 
tendency of dust to migrate away from the debris belt where it was 
created (Krivov et al. 2008), the presence of more or less narrow debris 
rings around a star has been attributed to the existence of unseen 
planet(s) shepherding the dust and confining its radial location through 
dynamical interaction, creating observable warps, clumps, gaps and 
asymmetries in the disc (Moro-Martin et al. 2007, Morales et al. 2009).  
Several such cases of planets interacting with a debris disc have now 
been proposed, with candidate planets identified through direct imaging 
searches around several of the stars (e.g. Vega, Wyatt 2003; Beta Pic, 
Lagrange et al. 2010; HD 95086, Rameau et al. 2013).  Indeed, 
planet-disc interaction has been vital in the formation and evolution of 
life on Earth, with minor body collisions providing both a late veneer 
of volatile material to the Earth's surface (O'Brien et al. 2006), the 
migration of Jupiter thought to be responsible for the late heavy 
bombardment at $\sim$800 Myr (Gomes et al. 2005) and subsequent 
infrequent catastrophic bombardment drastically altering the climate 
during the history of the solar system \citep{covey94, toon97, f09}.  
Therefore, any discussion of the potential habitability of an exoplanet 
should consider the possibility of volatile material delivery to a 
planet located in the habitable zone from remnant material located 
elsewhere in the system.

\acknowledgements

This research is supported by Australian Research Council grants 
DP0774000 and DP130102695.  Australian access to the Magellan Telescopes 
was supported through the National Collaborative Research Infrastructure 
Strategy of the Australian Federal Government.  This research has made 
use of NASA's Astrophysics Data System (ADS), and the SIMBAD database, 
operated at CDS, Strasbourg, France.  This research has also made use of 
the Exoplanet Orbit Database and the Exoplanet Data Explorer at 
exoplanets.org \citep{wright11}.



\begin{deluxetable}{lrr}
\tabletypesize{\scriptsize}
\tablecolumns{3}
\tablewidth{0pt}
\tablecaption{AAT/UCLES Radial Velocities for GJ 832}
\tablehead{
\colhead{JD-2400000} & \colhead{Velocity (\ms)} & \colhead{Uncertainty
(\ms)}}
\startdata
\label{AATvels}
51034.08733  &      7.5  &    2.2  \\
51119.01595  &     14.6  &    6.0  \\
51411.12220  &     11.4  &    3.3  \\
51683.26276  &     18.0  &    2.8  \\
51743.14564  &     19.0  &    2.7  \\
51767.08125  &     25.0  &    2.3  \\
52062.24434  &     19.8  &    2.2  \\
52092.16771  &      9.0  &    2.5  \\
52128.12730  &      2.2  &    4.0  \\
52455.23394  &      0.5  &    1.6  \\
52477.14549  &     10.0  &    2.6  \\
52859.08771  &     -4.1  &    2.1  \\
52943.03605  &     -5.4  &    2.7  \\
52946.97093  &      0.5  &    1.9  \\
53214.20683  &     -9.5  &    2.5  \\
53217.21195  &    -13.9  &    2.4  \\
53243.05806  &     -2.1  &    2.4  \\
53245.15092  &    -15.4  &    2.5  \\
53281.04691  &    -17.3  &    2.0  \\
53485.30090  &    -13.1  &    2.0  \\
53523.30055  &     -4.9  &    1.6  \\
53576.14194  &    -11.5  &    1.6  \\
53628.06985  &     -0.4  &    5.2  \\
53629.05458  &    -15.2  &    2.1  \\
53943.10723  &     -6.3  &    1.3  \\
54009.03770  &    -10.4  &    1.6  \\
54036.95562  &     -7.2  &    1.5  \\
54254.19997  &      3.2  &    1.8  \\
54371.06683  &      0.2  &    1.6  \\
54375.04476  &      2.5  &    1.7  \\
54552.29135  &      8.7  &    4.0  \\
54553.30430  &     17.0  &    2.8  \\
55102.99894  &      6.4  &    2.6  \\
55376.26506  &      9.2  &    2.5  \\
55430.16511  &     15.4  &    2.5  \\
56087.23879  &     16.1  &    2.4  \\
56139.24349  &     14.5  &    4.6  \\
56467.24320  &      1.6  &    3.0  \\
56499.09217  &     -6.3  &    4.0  \\
\enddata
\end{deluxetable}


\begin{deluxetable}{lrrr}
\tabletypesize{\scriptsize}
\tablecolumns{4}
\tablewidth{0pt}
\tablecaption{HARPS Radial Velocities for GJ 832}
\tablehead{
\colhead{JD-2400000} & \colhead{HARPS-TERRA} & \colhead{HARPS-CR} & 
\colhead{} \\ 
\colhead{} & \colhead{Velocity (\ms)} &\colhead{Velocity (\ms)} & 
\colhead{Uncertainty (\ms)}}
\startdata
\label{HARPSvels}
52985.51975  &     -5.1  & -8.4 &     0.5  \\
53158.90619  &     -3.1  & -6.6 &     0.6  \\
53205.74533  &     -5.4  & -9.0 &    0.7  \\
53217.74390  &     -8.7  & -12.4 &    0.4  \\
53218.70745  &     -9.3  & -12.8 &    0.2  \\
53229.72411  &     -6.7  & -10.2 &    0.4  \\
53342.54349  &     -9.6  & -13.3 &    0.6  \\
53490.92722  &    -12.1  & -16.0 &    0.8  \\
53491.91352  &     -7.7  & -11.3 &    1.3  \\
53492.92944  &     -8.4  & -15.1 &    1.0  \\
53551.85358  &     -7.8  & -11.7 &    0.3  \\
53573.80021  &    -11.2  & -15.1 &    0.3  \\
53574.73313  &    -12.1  & -16.9 &    0.3  \\
53575.73634  &    -10.4  & -14.5 &    0.3  \\
53576.78382  &    -11.1  & -16.0 &    0.3  \\
53577.79171  &    -11.0  & -15.1 &    0.3  \\
53578.74684  &    -11.9  & -15.9 &    0.3  \\
53579.72975  &    -11.3  & -14.5 &    0.3  \\
53580.76546  &    -10.4  & -13.8 &    0.3  \\
53950.81187  &     -8.6  & -12.9 &    0.4  \\
53974.63508  &     -6.6  & -11.0 &    0.4  \\
54055.52259  &     -4.0  & -8.4 &    0.4  \\
54227.91203  &     -1.2  & -5.7 &    0.4  \\
54228.91277  &     -1.2  & -3.9 &    0.4  \\
54230.88177  &     -0.3  & -3.4 &    0.4  \\
54233.92916  &     -0.1  & -3.9 &    0.7  \\
54234.92383  &      0.0  & -4.0 &    0.4  \\
54255.84319  &      1.7  & -2.9 &    0.4  \\
54257.88296  &      1.7  & -3.4 &    0.4  \\
54258.91849  &      0.8  & -3.9 &    0.4  \\
54291.81785  &      2.6  & -2.0 &   0.5  \\
54293.78153  &      3.4  & -1.6 &    0.3  \\
54295.82951  &      4.4  & -0.5 &    0.4  \\
54299.83521  &      6.9  & 2.0 &    0.7  \\
54314.77282  &      2.4  & -2.4 &    0.4  \\
54316.60452  &      1.0  & -4.5 &    0.2  \\
54319.80379  &      0.7  & -4.2 &    0.4  \\
54339.64829  &      5.2  & 0.5 &    0.2  \\
54341.76270  &      5.9  & 1.1 &    0.3  \\
54342.67095  &      3.0  & -2.7 &    0.1  \\
54347.71453  &      0.0  & -4.4 &   0.3  \\
54349.72920  &      0.8  & -2.6 &    0.4  \\
54387.61499  &      4.0  & -0.7 &    0.1  \\
54393.60450  &      3.6  & -1.3 &    0.4  \\
54420.51798  &      2.4  & -2.4 &    0.3  \\
54426.51760  &      3.4  & -1.0 &    0.3  \\
54446.53786  &      9.5  & 4.7 &    0.5  \\
54451.53099  &      9.5  & 4.5 &    0.5  \\
54453.53428  &      7.1  & 1.1 &    0.2  \\
54464.53844  &      6.2  & 1.3 &    0.5  \\
54639.91552  &     10.7  & 5.7 &    0.3  \\
54658.87484  &     15.7  & 10.7 &    0.6  \\
54662.86973  &     14.5  & 9.3 &    0.2  \\
54704.70377  &     11.6  & 6.5 &    0.5  \\
\enddata
\end{deluxetable}

\begin{deluxetable}{lrr}
\tabletypesize{\scriptsize}
\tablecolumns{3}
\tablewidth{0pt}
\tablecaption{Magellan/PFS Radial Velocities for GJ 832}
\tablehead{
\colhead{JD-2400000} & \colhead{Velocity (\ms)} & \colhead{Uncertainty
(\ms)}}
\startdata
\label{PFSvels}
55785.64157  &      0.0  &    0.9  \\
55787.61821  &      0.0  &    0.8  \\
55790.61508  &      0.3  &    0.8  \\
55793.63258  &      0.8  &    0.9  \\
55795.70095  &      1.2  &    0.8  \\
55796.71462  &      2.5  &    0.9  \\
55804.66221  &      0.2  &    0.9  \\
55844.60440  &      3.1  &    0.9  \\
55851.62322  &     -0.5  &    0.9  \\
56085.87962  &     -1.4  &    0.9  \\
56141.67188  &     -7.3  &    0.8  \\
56504.79755  &    -12.1  &    1.1  \\
56506.76826  &    -13.7  &    1.0  \\
56550.60574  &    -14.8  &    0.9  \\
56556.65127  &    -18.3  &    1.0  \\
56603.55010  &    -18.7  &    0.9  \\
\enddata
\end{deluxetable}

\begin{deluxetable}{lccc}
\tabletypesize{\scriptsize}
\tablecolumns{4}
\tablewidth{0pt}
\tablecaption{Maximum \textit{a posteriori} estimates and 99\% 
credibility intervals for the nuisance parameters: reference velocities 
with respect to the data mean ($\gamma$), excess white noise ($\sigma$), 
and intrinsic correlation ($\phi$). }
\tablehead{
\colhead{Parameter} & \colhead{HARPS} & \colhead{PFS} & \colhead{UCLES}
 }
\startdata
\label{instrument_parameters}
$\gamma$ & 4.56 [-1.83, 11.63] & -6.64 [-17.20, 3.91] & 1.14 [-5.52, 7.80] \\
$\sigma$ & 1.33 [0.90, 1.91] & 1.45 [0.44, 2.99] & 4.66 [3.13, 6.15] \\
$\phi$ & 0.90 [0.25, 1] & 0.77 [-1,1] & 0.11 [-0.30, 0.57] \\
\enddata
\end{deluxetable}


\begin{deluxetable}{lll}
\tabletypesize{\scriptsize}
\tablecolumns{3}
\tablewidth{0pt}
\tablecaption{Stellar Parameters for GJ 832 }
\tablehead{
\colhead{Parameter} & \colhead{Value} & \colhead{Reference}
 }
\startdata
\label{stellarparams}
Spec.~Type & M1.5 & \citet{gray06} \\
  & M1V & \citet{jenkins06} \\
Mass (\Msun) & 0.45 & \citet{bonfils13} \\ 
  & 0.45$\pm$0.05 & \citet{bailey09} \\
Distance (pc) & 4.95$\pm$0.03 & \citet{vl07} \\
log$R'_{HK}$ & -5.10 & \citet{jenkins06} \\
$[Fe/H]$ & -0.3$\pm$0.2 & \citet{bonfils13} \\
  & -0.7 & \citet{schiavon97} \\
$T_{eff}$ (K) & 3472 & \citet{casagrande08} \\
Luminosity (\Lsun) & 0.020 & \citet{boyajian12} \\
  & 0.026 & \citet{bonfils13} \\
log $g$ & 4.7 & \citet{schiavon97} \\
\enddata
\end{deluxetable}

\clearpage

\begin{deluxetable}{lcc}
\tabletypesize{\scriptsize}
\tablecolumns{3}
\tablewidth{0pt}
\tablecaption{Maximum \textit{a posteriori} estimates and 99\% 
credibility intervals of the Keplerian parameters and the linear trend 
$\dot{\gamma}$. }
\tablehead{
\colhead{Parameter} & \colhead{GJ\,832b} & \colhead{GJ\,832c}
 }
\startdata
\label{planetary_parameters}
$P$ [days] & 3660 [3400, 3970] & 35.67 [35.55, 35.82] \\
$K$ [ms$^{-1}$] & 15.51 [13.47, 17.36] & 1.62 [0.70, 2.56] \\
$e$ & 0.08 [0, 0.17] & 0.03 [0, 0.25] \\
$\omega$ [deg] & 246 [149, 304] & 80 [0, 360] \\
Mean anomaly\tablenotemark{a} [deg] & 40 [315, 109] & 246 [0, 360] \\
$a$ [AU] & 3.60 [3.18, 3.96] & 0.162 [0.145, 0.179] \\
$m \sin i$ [M$_{\oplus}$] & 219 [168, 270] & 5.0 [1.9, 8.1] \\
\hline
$\dot{\gamma}$ [ms$^{-1}$year$^{-1}$] & 0.18 [-0.46, 0.69] \\
$\gamma_{HARPS}$ & 4.56 [-1.83, 11.63] \\
$\gamma_{PFS}$ & -6.64 [-17.20, 3.91] \\
$\gamma_{AAT}$ & 1.14 [-5.52, 7.80] \\
\enddata
\tablenotetext{a}{Computed for epoch JD=2450000.0}
\end{deluxetable}



\begin{deluxetable}{lcc}
\tabletypesize{\scriptsize}
\tablecolumns{3}
\tablewidth{0pt}
\tablecaption{Correlations of GJ 832 velocities with activity indicators}
\tablehead{
\colhead{Indicator} & \colhead{MAP estimate} &
\colhead{99\% confidence interval} 
}
\startdata
\label{activity}
BIS & -0.21 & [-0.66,0.14] \\
FWHM & -0.04 & [-0.10,0.04] \\
S-index (m/s/dex) & 1.5 & [-8.2,11.0] \\
\enddata
\end{deluxetable}


\begin{deluxetable}{lcc}
\tabletypesize{\scriptsize}
\tablecolumns{3}
\tablewidth{0pt}
\tablecaption{Least-squares Keplerian orbital solutions for the GJ\,832 
planetary system. Uncertainties are given as a $\pm1\sigma$ range. }
\tablehead{
\colhead{Parameter} & \colhead{GJ\,832b} & \colhead{GJ\,832c}
 }
\startdata
\label{planetparams}
$P$ [days] & 3657 [3553, 3761] & 35.68 [35.65, 35.71] \\
$K$ [ms$^{-1}$] & 15.4 [14.7, 16.1] & 1.79 [1.52, 2.06] \\
$e$ & 0.08 [0.02, 0.10] & 0.18 [0.05, 0.31] \\
$\omega$ [deg] & 246 [224, 268] & 10 [323, 57] \\
Mean anomaly\tablenotemark{a} [deg] & 307 [285,330] & 165 [112,218] \\
$a$ [AU] & 3.56 [3.28, 3.84] & 0.163 [0.157, 0.169] \\
$m \sin i$ [M$_{\oplus}$] & 216 [188, 245] & 5.40 [4.45, 6.35] \\
\hline
$\dot{\gamma}$ [ms$^{-1}$year$^{-1}$] & 0.0 (fixed) \\
$\gamma_{HARPS}$ & 1.05 [0.27, 1.83] \\
$\gamma_{PFS}$ & -9.35 [-11.07, -7.63] \\
$\gamma_{AAT}$ & 3.08 [2.19, 3.96] \\
\enddata
\tablenotetext{a}{Computed for epoch JD=2450000.0}
\end{deluxetable}


\begin{deluxetable}{lcccccc}
\tabletypesize{\scriptsize}
\tablecolumns{7}
\tablewidth{0pt}
\tablecaption{Characteristics of 2-planet fits for GJ 832}
\tablehead{
\colhead{} & \colhead{AAT rms} & \colhead{HARPS-TERRA rms} & 
\colhead{HARPS-CR rms} & \colhead{PFS rms} & \colhead{Total rms} & 
\colhead{$\chi^2_{\nu}$}\\
\colhead{} & \colhead{\ms} & \colhead{\ms} & \colhead{\ms} & 
\colhead{\ms} & \colhead{\ms} & \colhead{}
}
\startdata
\label{rms}
Fit 1 & 5.66 & 1.53 & \nodata & 1.60 & 3.57 & 1.206 \\
Fit 2 & 5.63 & \nodata & 1.40 & 1.55 & 3.53 & 1.080 \\
\enddata
\end{deluxetable}

\begin{deluxetable}{llllllll}
\tabletypesize{\scriptsize}
\tablecolumns{8}
\tablewidth{0pt}
\tablecaption{FAP of residual signal after removing GJ\,832b }
\tablehead{
\colhead{Data Used} & \colhead{$N$} & \colhead{Period} & \colhead{FAP} & 
\colhead{Data Used} & \colhead{$N$} & \colhead{Period} & \colhead{FAP}\\
\colhead{} & \colhead{} & \colhead{(days)} & \colhead{} & \colhead{} & 
\colhead{} & \colhead{(days)} & \colhead{}
 }
\startdata
\label{faps}
AAT & 39 & 2.77 & 0.1481 & & & & \\
AAT + PFS & 55 & 15.38 & 0.7279 & & & & \\
HARPS-TERRA\tablenotemark{a} & 54 & 40.5 & 0.0066 & HARPS-CR\tablenotemark{a} & 54 & 35.6 & 0.0017 \\
HARPS-TERRA + PFS\tablenotemark{a} & 70 & 35.66 & 0.0331 & HARPS-CR + PFS\tablenotemark{a} & 70 & 35.66 & $<$0.0001 \\
AAT + HARPS-TERRA & 93 & 35.6 & 0.0461 & AAT + HARPS-CR & 93 & 35.7 & 0.0014 \\
AAT + HARPS-TERRA + PFS & 109 & 35.7 & 0.0164 & AAT + HARPS-CR + PFS & 109 & 35.7 & 0.0004 \\
\enddata
\tablenotetext{a}{Parameters of GJ\,832b held fixed at best-fit values 
in Table~\ref{planetparams}}
\end{deluxetable}

\begin{deluxetable}{llllll}
\tabletypesize{\scriptsize}
\tablecolumns{6}
\tablewidth{0pt}
\tablecaption{Candidate habitable-zone exoplanets }
\tablehead{
\colhead{Planet\tablenotemark{a}} & \colhead{Mass\tablenotemark{b} 
(\Mearth)} & \colhead{Semimajor axis (AU)} & \colhead{HZ range\tablenotemark{c} (AU)} & \colhead{Eccentricity} & 
\colhead{References\tablenotemark{d}}
 }
\startdata
\label{habitable}
GJ 163c & 6.8$\pm$0.9 & 0.1254$\pm$0.0001 & 0.134-0.237 & 0.099$\pm$0.086 & 1,2 \\
GJ 581g\tablenotemark{e} & 2.242$\pm$0.644 & 0.13386$\pm$0.00173 & 0.095-0.168 & 0.0 & 3,4,5,6,7,8,9 \\
GJ 581d & 5.94$\pm$1.05 & 0.21778$\pm$0.00198 & 0.095-0.168 & 0.0 & 8 \\
GJ 667Cc & 3.8 [2.6,6.3] & 0.125 [0.112,0.137] & 0.118-0.231 & 0.02 [0,0.17] & 10,11,12,13 \\  
GJ 667Cf & 2.7 [1.5,4.1] & 0.156 [0.139,0.170] & 0.118-0.231 & 0.03 [0,0.19] & \\
GJ 667Ce & 2.7 [1.3,4.3] & 0.213 [0.191,0.232] & 0.118-0.231 & 0.02 [0,0.24] & \\
GJ 832c & 5.406$\pm$0.954 & 0.163$\pm$0.006 & 0.130-0.237 & 0.18$\pm$0.13 & This work \\
HD 40307g & 7.1 [4.5,9.7] & 0.600 [0.567,0.634] & 0.476-0.863 & 0.29 [0,0.60] & 14 \\
Kepler-22b & $<$36 (1$\sigma$) & 0.849$^{+0.018}_{-0.017}$ & 0.858-1.524 & \nodata & 15,16 \\
Kepler-61b & \nodata & 0.2575$\pm$0.005 & 0.295-0.561 & 0.0$^{+0.25}_{-0.0}$ & 17,18 \\
Kepler-62e & $<$36 (95\%) & 0.427$\pm$0.004 & 0.457-0.833 & 0.13$\pm$0.112 & 17,19 \\
Kepler-62f & $<$35 (95\%) & 0.718$\pm$0.007 & 0.457-0.833 & 0.0944$\pm$0.021 & \\
Kepler-174d & \nodata & 0.677 & \nodata & 0.431-0.786 & 20 \\
Kepler-296f & \nodata & 0.263 & \nodata & 0.143-0.277 & 20 \\
Kepler-298d & \nodata & 0.305 & \nodata & 0.351-0.65 & 20 \\
Kepler-309c & \nodata & 0.401 & \nodata & 0.228-0.434 & 20 \\
\enddata
\tablenotetext{a}{Planet data from the Habitable Exoplanets Catalog at 
http://phl.upr.edu/hec.}
\tablenotetext{b}{Uncertainties given in square brackets refer to the 
99\% credibility intervals on the value in question, whilst those given 
as $\pm$ refer to the 1$\sigma$ uncertainty.}
\tablenotetext{c}{Conservative habitable-zone limits computed after \citet{kop14} and http://www3.geosc.psu.edu/~ruk15/planets/ }
\tablenotetext{d}{References in the table are as follows: [1] Bonfils et 
al. 2013, [2] Tuomi \& Anglada-Escud{\'e} 2014, [3] Mayor et al. 2009; [4] 
Vogt et al. 2010; [5] Tuomi 2011; [6] von Braun et al. 2011; [7] Tadeu 
dos Santos et al. 2012; [8] Vogt et al. 2012, [9] Lestrade et al. 2012; 
[10] Anglada-Escud{\'e} et al. 2012; [11] Anglada-Escud{\'e} et al. 2013; [12] 
Delfosse et al. 2013; [13] Makarov \& Berghea 2014; [14] Tuomi et al. 
2013a; [15] Borucki et al. 2012; [16] Neubauer et al. 2012; [17] Borucki 
et al. 2011; [18] Ballard et al. 2013; [19] Borucki et al. 2013; [20] 
Rowe et al. 2014; [21] Tuomi et al. 2013b.}
\tablenotetext{e}{For the purposes of this table, we list the cirular 
5-planet model for GJ\,581 given in Vogt et al. (2012).}
\end{deluxetable}


\end{document}